\documentclass[a4paper,12pt]{article}
\usepackage[english]{babel}
\usepackage[utf8]{inputenc}
\usepackage{siunitx}
\usepackage{authblk}
\usepackage{glossaries}
\setacronymstyle{long-short}
\usepackage{cite}
\usepackage{hyperref}
\usepackage{cleveref}
\usepackage{latexsym}
\usepackage{amsmath}
\usepackage{multicol}
\usepackage[pagewise]{lineno}
\usepackage{xpatch}

\makeatletter         
\def\@maketitle{

\begin{center}
{\Large \bfseries \sffamily \@title }\\[4ex] 
{\@author}\\[4ex] 
\@date\\[8ex]
\includegraphics[width = 40mm]{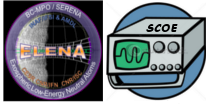}
\end{center}}
\makeatother

\usepackage{cleveref}
\usepackage{xcolor}
\usepackage[pagewise]{lineno}
\usepackage{fancyhdr}
\usepackage{lastpage}

\usepackage[ddmmyyyy]{datetime} 
\hypersetup{
    colorlinks=true,
    linkcolor=blue,
    filecolor=magenta,      
    urlcolor=cyan,
}
\newcounter{mybibstartvalue}
\xpatchcmd{\thebibliography}{%
  \usecounter{enumiv}%
}{%
  \usecounter{enumiv}%
  \setcounter{enumiv}{\value{mybibstartvalue}}%
}{}{}

\usepackage{graphicx}
\graphicspath{ {./pictures/} } 
\usepackage[font=small,labelfont=bf]{caption} 
\usepackage{booktabs} 
\usepackage{relsize} 
\DeclareUnicodeCharacter{00B0}{}
\usepackage{amsthm}
\theoremstyle{definition}
\fancypagestyle{plain}{%
  \fancyhf{}%
  \fancyhead[C]{\includegraphics[width = 13cm]{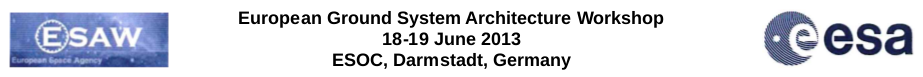} } %
}

\title{The BepiColombo SERENA/ELENA Instrument\\ 
On-Ground Testing with the ELENA\\ 
Special Check Out Equipment (SCOE)
}
\author[1]{Lazzarotto, F.}
\author[1]{Vertolli, N.}
\author[3]{Maschietti, D.}
\author[3]{Di Lellis A.M.}
\author[5]{Ryno, J.}
\author[6]{Camozzi, F.}
\author[1]{Orsini, S.}
\author[1]{Milillo, A.}
\author[1]{Mura, A.}
\author[1]{De Angelis, E.}
\author[1]{Rispoli, R.}
\author[1]{Colasanti, L.}
\author[2]{Selci S.}
\author[2]{D'Alessandro M.}
\author[4]{Leoni, R.}
\author[4]{Mattioli, F.}

\affil[1]{INAF-IAPS, Rome, Italy, \url{http://www.iaps.inaf.it/en/}} 
\affil[2]{ISM-CNR, Rome, Italy, \url{https://www.cnr.it/en/institute/087}}
\affil[3]{AMDL srl, Rome, Italy, \url{http://www.amdl.biz/}} 
\affil[4]{IFN-CNR, Rome, Italy, \url{http://www.roma.ifn.cnr.it/}}
\affil[5]{FMI, Helsinki, Finland, \url{https://en.ilmatieteenlaitos.fi/}} 
\affil[6]{OHB-Italia, Milan, Italy, \url{https://www.ohb-italia.it/}}
\date{June 21, 2013 - {\small mailto:francesco.lazzarotto@inaf.it}}  
\begin{document}

\maketitle

\begin{abstract}
The neutral particles sensor ELENA\cite{10.1117/1.OE.52.5.051206} (Emitted Low-Energy Neutral Atoms) for the ESA/JAXA 
BepiColombo\cite{BC} mission to Mercury\cite{MM} (in the SERENA\cite{2010P&SS...58..166O}
instrument suite) is devoted to measure  low energetic neutral atoms. 
The main goal of the experiment is measuring the sputtering emission from planetary surfaces, 
from $E=~20eV \enspace$up to$\enspace E=~5keV$, 
within 1-D ($2\si{\degree} \times 76\si{\degree}$). ELENA original project had also a particle 
discrimination system based on Time-of-Flight (TOF) of particles through the shutter on the 
Micro Channel Plates detector (MCP), it has been withdrawn from the flight model 
due to design and development problems. 
The ELENA SCOE is the configuration/testing system of ELENA, it allows to command operations and to set up 
configuration parameters on the instrument and to monitor the incoming data. 
The TC/TM simulation/encoding/decoding software is developed respecting 
the CCSDS/ECSS standards implemented by ESA, and it's SCOS2000\cite{SCOS2000} compatible. 
TC generation, HK data monitoring and basic science data analysis are operated by the SERENA EGSE, 
developed by the Finnish Meteorological Institute (\href{https://en.ilmatieteenlaitos.fi/}{FMI}), Helsinki, Finland. 
The data stream outcoming from the EGSE is then preprocessed from TM to user readable formats: FITS (\url{http://fits.gsfc.nasa.gov/}) 
and then ASCII csv tables with metadata collected in a detached XML file, called \textit{label}. 
This task is performed using the PacketLib, ProcessorLib, and DISCoS (PPD) framework (see \cite{2003ASPC..295..473B}) 
and is going to be used as the first level prototype of the BepiColombo Science Ground Segment processing pipeline, based in ESAC, Madrid, Spain 
and implemented using the PDS4 data format (\url{http://pds.nasa.gov/pds4/about/what.shtml}).
\end{abstract}
\section{The ELENA instrument data production chains}
The system developed in cooperation by FMI, INAF  and industrial partners is able to assure 
data monitoring/transmission functions for all the ELENA data production modes, 
the first ELENA data type managed by the teams will be the basic mode 
(histogram of angular sectors). ELENA data TC/TM packets are described in the document
BC-SRN-RS-31000 "SERENA ELENA S/W Specification`` (by OHB \& AMDL). 
The ELENA data production chains are the following:
\begin{enumerate}
\item  full, event-by-event: array of events, 1 event = 16 bits;
\item  basic, h16: histogram of the sectors (16 channels);
\item  basic, h32: histogram of the sectors (32 channels).
\end{enumerate}

\section{The TC/TM Packet Viewer}
\begin{center}
\includegraphics[width=0.95\linewidth]{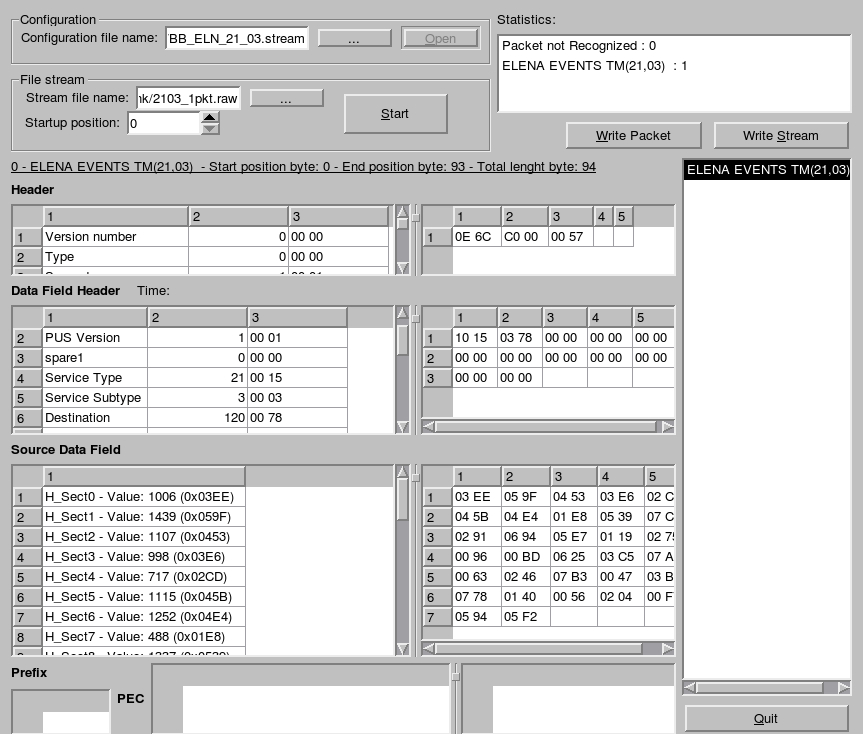}
\captionof{figure}{The TM Packet Viewer displaying ELENA data}
\end{center}
It's a tool able to show the data contained in the
SERENA/ELENA TM packets. It allows engineering
monitoring functionalities (parameter control, debug), with
a proper configuration it is able to decode and handle any kinds of TC/TM
data packets respecting the international CCSDS and european ECSS standards
\section{The SERENA EGSE}
\begin{center}
\includegraphics[width=1\linewidth]{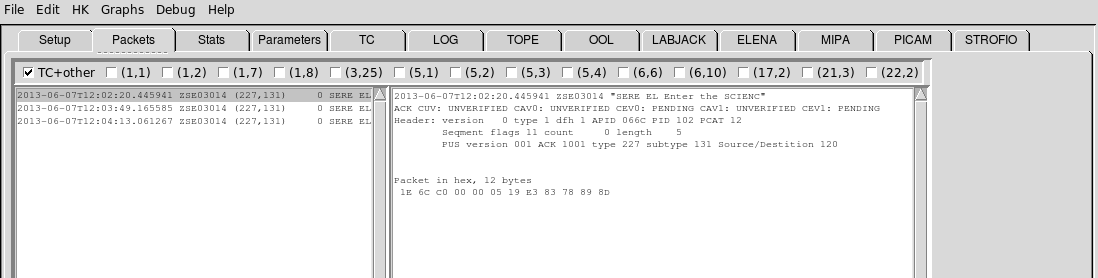}
\captionof{figure}{ELENA TM Packets received and displayed by SERENA EGSE}
\end{center}
The ELENA TM packets are read using the SERENA EGSE developed by FMI (see \url{http://www.erlang.se/euc/04/ryno.pdf} 
for details). The data stream is received from spacewire and serial interfaces, then redirected to a TCP connection 
and saved on binary files. TCs are generated using DB definition and received by the prototypes through spacewire, 
serial and TCP links.

\section{TM Preprocessing and Display}
\begin{center}
\includegraphics[width=0.8\linewidth]{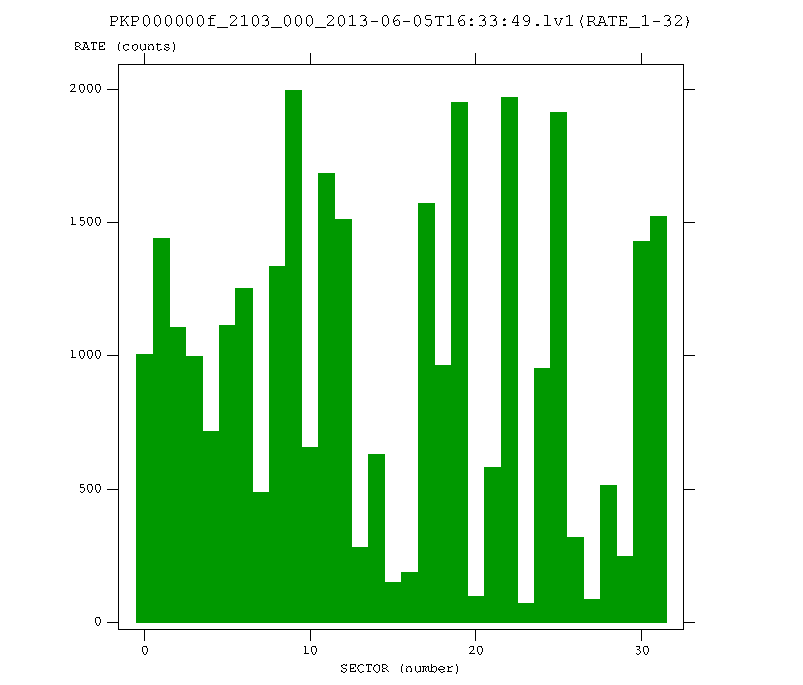}
\captionof{figure}{Display of the histogram data preprocessed into FITS binary tables}
\end{center}
The incoming TM data stream is converted in user readable
format data files (ASCII/csv, FITS and XML) using the PacketLib,
ProcessorLib, DISCoS framework (see also \cite{2001ASPC..238..245G} and \cite{TrifoglioTMPPS-SPIE2008} ), 
creating C++ programs wrapoed by Bash scripts and text configuration files
describing the packet structure for each data production mode. 
Data production software simulators are created and used in order to
develop the Ground System in parallel to the construction of the instrument prototypes.
\section{References}
\renewcommand{\section}[2]{\vskip 0.05em} 
{
\small
\bibliography{fralaz1971_ESAW2013}{}

\begin{thebibliography}{1}

\bibitem{BC}
European~Space Agency.
\newblock {BepiColombo} science mission homepage.
\newblock \url{https://sci.esa.int/web/bepicolombo}.

\bibitem{2003ASPC..295..473B}
A.~{Bulgarelli} et~al.
\newblock {\em {PacketLib: A C++ Library for Scientific Satellite Telemetry
  Applications}}, volume 295 of {\em Astronomical Society of the Pacific
  Conference Series}, page 473.
\newblock 2003.

\bibitem{2001ASPC..238..245G}
F.~{Gianotti} and M.~{Trifoglio}.
\newblock {\em {DISCoS---Detector-Independent Software for On-Ground Testing
  and Calibration of Scientific Payloads Using the ESA Packet Telemetry and
  Telecommand Standards}}, volume 238 of {\em Astronomical Society of the
  Pacific Conference Series}, page 245.
\newblock 2001.

\bibitem{2010P&SS...58..166O}
S.~{Orsini}, S.~{Livi}, K.~{Torkar}, S.~{Barabash}, and the {SERENA Team}.
\newblock "{SERENA}: A suite of four instruments (elena, strofio, picam and
  mipa) on board bepicolombo-mpo for particle detection in the hermean
  environment".
\newblock {\em Planetary and Space Science}, 58(1-2):166--181, Jan 2010.

\bibitem{SCOS2000}
M.~{Pecchioli} and the ESA Operation Centre System Infrastructure Section
  (OPS-GIC).
\newblock "scos-2000 database import interface control document".
\newblock
  \url{http://emits.sso.esa.int/emits-doc/Annex-J_S3-EGOS-MCS-S2K-ICD-0001_I6.2_SCOS2000-DI-ICD.pdf},
  2006.

\bibitem{10.1117/1.OE.52.5.051206}
Rosanna Rispoli et~al.
\newblock "{ELENA} microchannel plate detector: absolute detection efficiency
  for low energy neutral atoms".
\newblock {\em Optical Engineering}, 52(5):1 -- 8, 2013.

\bibitem{TrifoglioTMPPS-SPIE2008}
M.~{Trifoglio} et~al.
\newblock {\em Architecture and performances of the AGILE Telemetry
  Preprocessing System (TMPPS)}, volume 7011 , 70113E of {\em Proc. of SPIE},
  pages 3E 1--8.
\newblock 2008.

\bibitem{MM}
P.~{Wurz}.
\newblock "{Misterious Mercury}" (in spatium n.29).
\newblock \url{http://www.issibern.ch/publications/pdf/spatium/Spatium_29.pdf},
  2012.

\end{thebibliography}
\bibliographystyle{plain}
}

\end{document}